\documentclass[aps,twocolumn,letter,showpacs]{revtex4}
\usepackage{epsfig}
\newcounter{list1} 
\begin{document}
\title{Extreme times in financial markets}
\author{Jaume Masoliver, Miquel Montero, Josep Perell\'o,}
\affiliation{Departament de F\'{\i}sica Fonamental, Universitat de Barcelona,\\ Diagonal, 647, E-08028 Barcelona, Spain}
\date{\today}

\begin{abstract}
We apply the theory of continuous time random walks to study some aspects of the extreme value problem applied to financial time series. We focus our attention on extreme times, specifically the mean exit time and the mean first-passage time. We set the general equations for these extremes and evaluate the mean exit time for actual data.
\end{abstract}
\maketitle

\section{Introduction}

The study of financial market microstructure has been the object of intense research since the beginning of mathematical finance \cite{ohara,lo}. Thus one of the first models of market microstructure was set by Bachelier who proposed the mechanism of the random walk (RW) to represent the dynamics of speculative prices \cite{cootner}. In fact Bachelier's RW, with slight modifications and additions, has been successfully applied to a variety of problems in finance ranging from price dynamics to option pricing \cite{lo,hull}. Unfortunately, RW models because of their Gaussian nature fail to account for a universal characteristic of price distributions: the fact that they are highly leptokurtic, with considerable fat tails at the wings of the distributions 
\cite{mantegnastanley,gabaix}. This was one of the main reasons which led Mandelbrot to propose the Levy distribution as an alternative description of the probability distribution of prices \cite{mandelbrot}. However, pure Levy models have a serious drawback since any statistical moment, beyond the first one, does not exist. This fact has induced some authors to propose truncated Levy walks (TLW) as an alternative model. TLW's have the advantage of having finite moments up to some order (which depends on the truncation procedure) while keeping, to some extend, some quite attractive properties of TL's such as self-similarity and scaling laws \cite{shlesinger,plerou}. One of the problems of TL's is that the truncation is an ad-hoc procedure with little relation, if any, with the underlying dynamics. Alternative models based on the Gaussian process plus jumps \cite{merton} or even continuous jumps \cite{mmp} have also been proposed with acceptable results.

Continuous Time Random Walks (CTRW's) are general models which perhaps better capture market microstructure, specially that of high frequency data. CTRW's were first introduced by Montroll and Weiss in 1965 \cite{montrollweiss} and have a long history of successful applications to physics, chemistry and geophysics to name a few \cite{havlin,weissllibre}.  To our knowledge, the application of the CTRW to finance is quite recent and it has not been fully developed its potential \cite{scalas,kutner,mmw,mmpw}. One of the applications where CTRW's may represent a valuable achievement is in the field of risk control because the CTRW formalism provides a natural way of treating any particular realization of the price or return processes. This, in turn, facilitates the study of extreme events which are central to risk management. 

The statistics of extremes is a difficult field in probability theory and its thorough description for a given random process can be quite involved, if not impossible, from an analytical point of view \cite{weissrubin,gardiner,katja}. In this paper we will use the CTRW formalism to study some aspects of the extreme value problem applied to finance. We will focus our attention on two of the simplest quantities related to extreme statistics: the mean first-passage time (MFPT) and the mean exit time (MET). MFPT is the average time at which the random process reaches, for the first time, some pre-assigned value while MET is the mean time when the random process leaves, for the first time, a given interval. 

The paper is organized as follows. In Sec. II we outline the CTRW formalism and briefly describe some of its applications. In Sec. III we develop the theory of MET for financial time series and apply it to real data in Sec. IV. Closing remarks are in Sec. V.

\section{Outline of the CTRW}

In this section we summarize the main features of the CTRW formalism applied to the analysis of financial time series. We refer to the reader to \cite{mmw,mmpw} for a more complete account on the subject. 

Let $S(t)$ be a financial price and let $t_0$ be any initial time. The log-price or return is defined by \mbox{$Z(t)=\ln S(t)/S(t_0)$}. If $\langle Z(t)\rangle$ is the return mean value, we define the zero-mean return by
\begin{equation}
X(t)=Z(t)-\langle Z(t)\rangle.
\label{x}
\end{equation}
For the rest of the paper we will assume that the return is a stationary random process independent of any initial time $t_0$ which is taken to be zero, $t_0=0$. Hence $t$ refers to a time interval. 

\begin{figure}
\epsfig{file=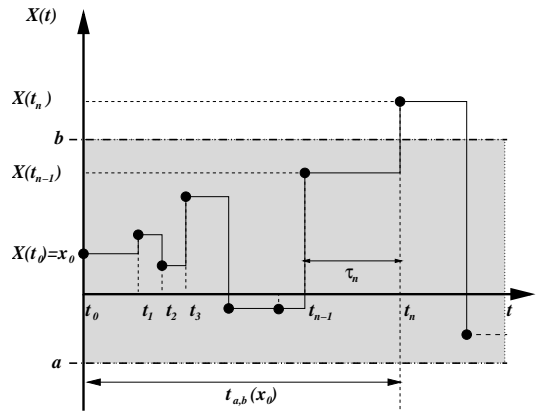, width=7cm} \caption{A particular trajectory of the zero-mean return process along with a particular value of the random variable $t_{a,b}(x_0)$.} \label{fig1}
\end{figure}

Suppose that $X(t)$ is described by a CTRW. In this representation any trajectory consists of a series of step functions as shown in Fig. \ref{fig1}. Therefore, $X(t)$ changes at random times $t_0,t_1,t_2,\cdots,t_n,\cdots$ and we assume that sojourns or waiting times, $\tau_n=t_n-t_{n-1}$ ($n=1,2,\cdots,n$), are independent and identically distributed random variables described by a given probability density function (pdf) defined by $\psi(t)dt=\mbox{Prob}\{t<\tau_n\leq t+dt\}$. At the conclusion of a given sojourn the return $X(t)$ suffers a random jump described by the random variable $\Delta X_n=X(t_n)-X(t_{n-1})$ whose pdf is defined by $h(x)dx=\mbox{Prob}\{x<\Delta X_n\leq x+dx\}$. We combine these two causes of randomness into one single density $\rho(x,t)$ which represents the joint pdf of waiting times and random jumps, {\it i.e.}, 
$$
\rho(x,t)dxdt=\mbox{Prob}\{x<\Delta X_n\leq x+dx;t<\tau_n\leq t+dt\}.
$$
We will further assume that $\rho(x,t)$ is an even function of $x$ so that there is no net drift in the evolution of $X(t)$. Note that if waiting times and jumps are independent random quantities, then $\rho(x,t)=h(x)\psi(t)$. In any other situation one has to specify a functional form of $\rho(x,t)$ that is compatible with the observed data. Moreover, since the jump pdf, $h(x)$, and the waiting-time pdf, $\psi(t)$, are the marginal densities of the joint density, any proposed form of $\rho(x,t)$ must satisfy
\begin{equation}
\psi(t)=\int_{-\infty}^{\infty}\rho(x,t)dx;\qquad h(x)=\int_{0}^{\infty}\rho(x,t)dt.
\label{marginals}
\end{equation}

The main objective of the CTRW is obtaining the so-called propagator, that is, the probability density function of the zero-mean return $X(t)$:
\begin{equation}
p(x,t)dx=\mbox{Prob}\{x<X(t)\leq x+dx\}.
\label{p(x,t)}
\end{equation}
In \cite{mmw,mmpw} we have obtained a general expression for the joint Fourier-Laplace transform of the propagator,
$$
\hat{p}(\omega,s)=\int_0^\infty e^{-st} dt\int_{-\infty}^{\infty} e^{-i\omega x} p(x,t)dx,
$$
in terms of the Laplace transform of the waiting time distribution, $\hat{\psi}(s)$, and the Fourier-Laplace transform of the joint distribution, $\hat{\rho}(\omega,s)$. This expression reads
\begin{equation}
\hat{p}(\omega,s)=\frac{[1-\hat{\psi}(s)]/s}{1-\hat{\rho}(\omega,s)}.
\label{p(omega,s)}
\end{equation}

Equation (\ref{p(omega,s)}) constitutes the most general solution to the problem. In order to proceed further we would need to specify the form of the joint density $\rho(x,t)$ which has to be guessed from data \cite{footnote0}. There are nonetheless some general results that are independent of the model chosen for $\rho(x,t)$~\cite{mmpw}. We briefly outline them:

\begin{list}
{(\alph{list1})}{\usecounter{list1}}
\item 
If the mean waiting time $\langle\tau\rangle$ is finite and the jump pdf $h(x)$ has a finite second moment, \mbox{$\mu_2=\langle \Delta X^2\rangle<\infty$}, the asymptotic distribution of returns for long times approaches to the Gaussian density:
\begin{equation}
\tilde{p}(\omega,t)\simeq e^{-\mu_2\omega^2t/2\langle\tau\rangle} \qquad (t\gg\langle\tau\rangle),
\label{gaussian}
\end{equation}
and the asymptotic variance grows linearly with time:
\begin{equation}
\langle X^2(t)\rangle\simeq\frac{\mu_2}{\langle\tau\rangle}\ t \qquad (t\gg \langle\tau\rangle).
\label{asympvol}
\end{equation}

\item 
If $h(x)$ is a long-tailed density, {\it i. e.,} $h(x)\sim|x|^{-1-\alpha}$ as $|x|\rightarrow\infty$. Then $\tilde{h}(\omega)\simeq 1-k|\omega|^{\alpha}$ as $\omega\rightarrow 0$ for $0<\alpha<2$. Moreover, if we assume that for $\omega$ small $\langle\tau e^{i\omega\Delta X}\rangle\simeq\langle\tau \rangle$, then the asymptotic return pdf approaches to the L\'evy distribution:
\begin{equation}
\tilde{p}(\omega,t)\simeq e^{-k|\omega|^\alpha t/\langle\tau \rangle} \qquad (t\gg\langle\tau \rangle).
\label{levy}
\end{equation}

\item
At intermediate times, $t\approx\langle\tau \rangle$, the behavior of $p(x,t)$ for large values of $|x|$, is the same as that of the jump distribution: 
\begin{equation}
p(x,t)\sim \frac{t}{\langle\tau \rangle}h(x).
\label{tailsp}
\end{equation}
\end{list}

From points (a) and (b) we see that the CTRW formalism unifies two of the approaches mentioned in Sec. I: the Bachelier's RW (Gaussian models) and Mandelbrot's Levy models, thus providing a natural way of understanding the truncation procedures appearing in TLW  models. Finally, point (c) shows that fat tails in price distributions have their origin in fat tails in the jump distribution. Hence, by observing the jump pdf $h(x)$ which is more easily accessible than the complete pdf $p(x,t)$, one can infer the approximate behavior at the tails of $p(x,t)$ via Eq. (\ref{tailsp}). We therefore see that {\it the CTRW formalism unifies in a single model the main characteristics observed in the empirical distributions of prices}.

\section{Extreme times}

We assume that the zero-mean return process $X(t)$ is described by a CTRW with a given waiting-time distribution, $\psi(t)$, jump distribution, $h(x)$, and joint density, $\rho(x,t)$. Suppose that initially the return has a known value $X(0)=x_0$ \cite{footnote1}. For $t>0$ we ask ourselves the following question: at which time interval $X(t)$ leaves a given interval $[a,b]$ for the first time? In other words, at which time the return is greater than certain value $b$ or smaller than $a$ for the first time? We call this quantity the exit time out of the interval $[a,b]$ and denote it by $t_{a,b}(x_0)$. Obviously $t_{a,b}(x_0)$ is a random variable since it depends on the particular trajectory of $X(t)$ chosen (see Fig. \ref{fig1}). Our main objective here is to obtain, based on the CTRW formalism, the mean exit time (MET) $T_{a,b}(x_0)=\langle t_{a,b}(x_0)\rangle$ 
\cite{footnote2}. 

The standard approach to MET problems requires the knowledge of the survival probability of the process in the interval $[a,b]$ \cite{weissrubin,gardiner,katja}. Although the interest in knowing the survival probability is beyond any doubt, its attainment turns out to be quite involved. In this paper we present a direct and much simpler derivation of the MET and leave obtaining the survival probability for a later work. 

We decompose $T(x_0)$ in two summands
\begin{equation}
T(x_0)=T_1(x_0)+T_2(x_0),
\label{t1+t2}
\end{equation}
where $T_1(x_0)$ is the MET to leave the interval $[a,b]$ in {\it only one jump} and $T_2(x_0)$ is the MET when more jumps have occurred. Note that in terms of the joint density $\rho(x,t)$, $T_1(x_0)$ can be written as
\begin{eqnarray}
T_1(x_0)&=&\int_0^{\infty}tdt\int_b^{\infty}\rho(x-x_0,t)dx\nonumber\\
&+&
\int_0^{\infty}tdt\int_{-\infty}^a\rho(x-x_0,t)dx,
\label{t1}
\end{eqnarray}
where the first summand is the mean time for the random walker to escape through the upper boundary $x=b$, and the second summand is the mean time to escape through the lower boundary $x=a$. If the random walker has not exited the interval in the first jump the process will have attained at time $t$ some value $x\in[a,b]$ inside the interval and from that point the mean exit time will be exactly $T(x)$. That is,
\begin{equation}
T_2(x_0)=\int_0^{\infty}dt\int_a^b\rho(x-x_0,t)[t+T(x)]dx.
\label{t2}
\end{equation}
Substituting Eqs. (\ref{t1})-(\ref{t2}) into Eq. (\ref{t1+t2}) and some simple algebra involving the use of Eq. (\ref{marginals}) finally yield the following integral equation for the MET $T(x_0)$
\begin{equation}
T(x_0)=\langle\tau\rangle+\int_a^bh(x-x_0)T(x)dx,
\label{met}
\end{equation}
where $\langle\tau\rangle$ is the mean waiting time between jumps:
\begin{equation}
\langle\tau\rangle=\int_0^{\infty}t\psi(t)dt.
\label{tau}
\end{equation}

From a mathematical point of view Eq. (\ref{met}) is a Fredholm integral equations of second kind. Depending on the specific nature of the kernel $h(x)$ there are some analytical approaches which allow to get an exact solution \cite{tricomi}. In the most general case if the kernel norm defined by 
\begin{equation}
||h||^2=\int_{a}^{b}\int_{a}^{b}h^2(x-y)dxdy,
\label{norm}
\end{equation}
is finite there is always a series solution that in many situations can be useful to obtain a good approximation \cite{tricomi}. In the next section we will see examples of exact and approximate solutions. 

An important point should be emphasized: the fact that, as shown in Eq. (\ref{met}), the MET does not depend on the possible coupling between waiting times and jumps. In other words, $T(x_0)$ is independent of the particular form of the joint density $\rho(x,t)$. 

Another extreme time closely related to the MET is the mean first-passage time, MFPT, defined as the mean time at which the process attains a given value $x_c$ for the first time. Let $T_{a,b}(x_0)$ be the MET out of the interval $[a,b]$ then if $x_0<x_c$ the MFPT to $x_c$ is defined as $T_c(x_0)=\lim_{a\rightarrow-\infty}T_{a,x_c}(x_0)$, while if $x_0>x_c$ we have $T_c(x_0)=\lim_{b\rightarrow\infty}T_{x_c,b}(x_0)$. Thus we see from Eq. (\ref{met}) that, if $x_0<x_c$, the MFPT obeys the integral equation
\begin{equation}
T_c(x_0)=\langle\tau\rangle+\int_{-\infty}^{x_c}h(x-x_0)T_c(x)dx,
\label{mfpt}
\end{equation}
with an analogous equation when $x_0>x_c$. Unfortunately, Eq. (\ref{mfpt}) is a singular integral equation for which the kernel norm defined above is infinite \cite{footnote3}. In such a case the solution of Eq. (\ref{mfpt}) may not exist \cite{tricomi}. Again, we will see an example of this in the next section.

\section{Some properties and results}

Equation (\ref{met}) constitutes the main result of this paper. In this section we will summarize some of the general properties of Eq. (\ref{met}) and present some results. We stress the fact, already mentioned in the previous section, that $T(x_0)$ depends on the entire jump distribution, $h(x)$, but does not depend on the whole waiting-time pdf, $\psi(t)$, being only necessary for the existence of the MET that the mean waiting time $\langle\tau\rangle$ be finite. We also observe that Eq. (\ref{met}) is written for a general and non-symmetrical interval $[a,b]$. However, we can always transform the problem in order to work on a symmetrical interval, something that lightens the problem. Thus we can easily see that the symmetrical MET defined by 
\begin{equation}
T_{\footnotesize\mbox{sym}}(x)\equiv T\left(x+\frac{a+b}{2}\right),
\label{tsim}
\end{equation}
satisfies the equation
\begin{equation}
T_{\footnotesize\mbox{sym}}(x_0)=\langle\tau\rangle+
\int_{-L/2}^{L/2}h(x-x_0)T_{\footnotesize\mbox{sym}}(x)dx,
\label{metsim}
\end{equation}
where $L=b-a$ is the length of the interval. When the jump distribution, $h(x)=h(-x)$, is even we see from Eq. (\ref{metsim}) that 
$T_{\footnotesize\mbox{sym}}(-x)$ satisfies the same equation as $T_{\footnotesize\mbox{sym}}(x)$. In other words, $T_{\footnotesize\mbox{sym}}(x)=T_{\footnotesize\mbox{sym}}(-x)$ is also an even function which implies that the derivative of $T_{\footnotesize\mbox{sym}}(x)$ at $x=0$ is zero if it exists, $T'_{\footnotesize\mbox{sym}}(0)=0$. 

There are few cases in which it is possible to get an exact solution for the MET problem. This is the case when the jump pdf is given by the Laplace distribution
\begin{equation}
h(x)=\frac{\gamma}{2}e^{-\gamma|x|},
\label{laplace}
\end{equation}
where $\gamma>0$ and $\langle\Delta X^2\rangle=2/\gamma^2$ is the variance of jumps. One can show that in this case the integral equation (\ref{met}) is equivalent to the following differential equation
\begin{equation}
T''(x_0)=-\gamma^2\langle\tau\rangle,
\label{de}
\end{equation}
with boundary conditions
\begin{equation}
T'(a)=\gamma[T(a)-\langle\tau\rangle], \quad T'(b)=-\gamma[T(b)-\langle\tau\rangle]. 
\label{bc}
\end{equation}
The solution to this problem reads 
\begin{equation}
T(x_0)=\frac{\langle\tau\rangle}{2}\left[1+\left(1+\frac{\gamma L}{2}\right)^2-\gamma^2\left(x_0-\frac{a+b}{2}\right)^2\right].
\label{exactmet}
\end{equation}

Note that in this case the MFPT is infinite since $T(x_0)\rightarrow\infty$, both as $a\rightarrow-\infty$ and $b\rightarrow\infty$. Consequently $T_c(x_0)=\infty$. We also note that the MET is a quadratic function of the interval length $L=b-a$. This is even more clearly seen by assuming a symmetrical interval $b=-a=L/2$ and that the initial return is zero, $x_0=0$ (in fact, this is the usual situation \cite{footnote1}). 
In this case Eq. (\ref{exactmet}) reads
\begin{equation}
T(0)=\frac{\langle\tau\rangle}{2}\left[1+\left(1+\frac{\gamma L}{2}\right)^2\right].
\label{exactmetsim}
\end{equation}

It is very illustrative to compare the above expressions for the MET with those of the ordinary random walk. If the price process follows a RW then in the continuous limit the zero-mean return is the Wiener process, {\it i.e.}, $X(t)=\sigma W(t)$, where $\sigma$ is the volatility. In this case the MET out of $[a,b]$ is \cite{gardiner}
\begin{equation}
T_{\mbox{\tiny RW}}(x_0)=\frac{1}{\sigma^2}(x_0-a)(b-x_0).
\label{metrw}
\end{equation}
Observe that boundary conditions now are $T_{\mbox{\tiny RW}}(a)=T_{\mbox{\tiny RW}}(b)=0$ which are quite different than those give by Eq. (\ref{bc}). In order to compare this time with that of the CTRW just obtained in Eq. (\ref{exactmet}) we will scale both times. We thus define the following dimensionless MET's:
$$
T^*_{\mbox{\tiny RW}}(x_0)=(\sigma^2/2L^2)T_{\mbox{\tiny RW}}(x_0) 
$$
and 
$$
T^*_{\mbox{\tiny CTRW}}(x_0)=
(1/\langle\tau\rangle\gamma^2L^2)T_{\mbox{\tiny CTRW}}(x_0).
$$
Then for a symmetrical interval $b=-a=L/2$ we get from Eqs. (\ref{exactmet}) and 
(\ref{metrw}) 
\begin{equation}
T^*_{\mbox{\tiny CTRW}}(x_0)=T^*_{\mbox{\tiny RW}}(x_0)+\frac{1}{\gamma^2L^2}+\frac{1}{2\gamma L},
\label{compare}
\end{equation}
where
\begin{equation}
T^*_{\mbox{\tiny RW}}(x_0)=\frac{1}{8}-\frac{x_0}{2L^2}.
\label{metrw2}
\end{equation}
We see that $T^*_{\mbox{\tiny CTRW}}(x_0)\rightarrow T^*_{\mbox{\tiny RW}}(x_0)$ when $L\gg\gamma^{-1}$, that is, when the length of the interval is much larger than the jump volatility which is proportional to $1/\gamma$. In this case our CTRW approaches the Wiener process with a volatility given by $\sigma=\sqrt{2}/\gamma\langle\tau\rangle$.

Let us now obtain an approximate solution which will be valid for any sufficiently smooth kernel $h(x)$. Suppose we have an even and zero-mean jump density satisfying the following scaling condition
\begin{equation}
h(x)=\frac{1}{\kappa}H\left(\frac{x}{\kappa}\right),
\label{scaling}
\end{equation}
where $\kappa>0$ is the volatility of the jump process (note that the density (\ref{laplace}) satisfies this condition with $\kappa=\sqrt{2}/\gamma$). In this case 
Eq. (\ref{metsim}) can be written as
\begin{equation}
\bar{T}(u)=\langle\tau\rangle+\int_{-L/2\kappa}^{L/2\kappa}H(v-u)\bar{T}(v)dv,
\label{metsimmod}
\end{equation}
where $\bar{T}(u)\equiv T_{\footnotesize\mbox{sym}}(\kappa u)$ and $-L/2\kappa\leq u\leq L/2\kappa$. Once we have an expression for $\bar{T}(u)$ the mean exit time is given, via Eq. (\ref{tsim}), by
\begin{equation}
T(x_0)=\bar{T}\left(\frac{2x_0-a-b}{2\kappa}\right).
\label{bart}
\end{equation}
Suppose now that $\epsilon\equiv L/2\kappa$ is small. In this case an approximate solution to Eq. (\ref{metsimmod}) can be easily obtained through an iteration procedure with the final result up to second order \cite{footnote4}:
\begin{eqnarray*}
\bar{T}(u)=\langle\tau\rangle\bigl[1+2H(0)\epsilon&+&(H'(0+)+4H(0)^2)\epsilon^2\\
&+& H'(0+)u^2+O(\epsilon^3)\bigr].
\end{eqnarray*}
Hence
\begin{eqnarray}
T(x_0)&=&\langle\tau\rangle\Biggl[1+H(0)\left(\frac{L}{\kappa}\right)+
[H'(0+)/4+H(0)^2]\left(\frac{L}{\kappa}\right)^2\nonumber\\
&+&
\frac{H'(0+)}{4\kappa^2}(2x_0-a-b)^2+O\left(\frac{L^3}{\kappa^3}\right)\Biggr].
\label{metapprox}
\end{eqnarray}
In the symmetrical case with $x_0=0$ we have
\begin{eqnarray}
T(0)=\langle\tau\rangle\Biggl[1+H(0)\left(\frac{L}{\kappa}\right) 
&+&
[H'(0+)/4+H(0)^2]\left(\frac{L}{\kappa}\right)^2 \nonumber\\
&+&O\left(\frac{L^3}{\kappa^3}\right)\Biggr].
\label{metapproxsym}
\end{eqnarray}
We see from these expressions that, in general, the MET has for sufficiently small intervals a quadratic growth behavior as in the case of the exponential density (\ref{laplace}). In fact, in this latter case \mbox{$\kappa=\sqrt{2}/\gamma$}, \mbox{$H(x)=\exp[-\sqrt{2}|x|]/\sqrt{2}$} and Eq.~(\ref{metapprox}) agrees with Eq.~(\ref{exactmet}). In other words the approximate expression becomes the exact solution. 

\begin{figure}
\epsfig{file=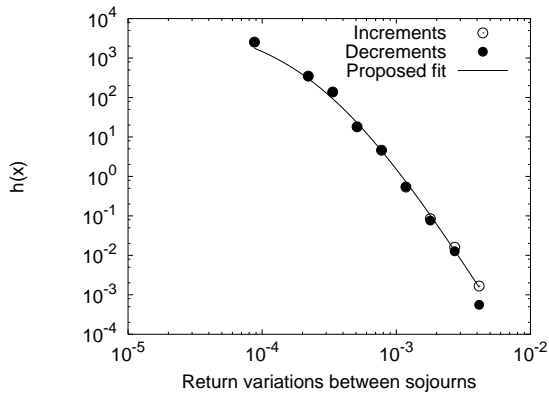, width=7.5cm}
\caption{Empirical distribution of the jump distribution $h(x)$ for the U.S. dollar/Deutsche mark futures market. This figure suggests the presence of a power law which is adjusted by Eq.~(\ref{empiricalh}) with the parameters reported in the main text.} 
\label{fig2}
\end{figure}

Let us finally apply the above results to real data. The data consist of tick-by-tick prices for the U.S. dollar/Deutsche mark future exchange from January 1993 to December 1997 (a total of 1,048,590 data points). The empirical jump distribution is plotted in 
Fig. \ref{fig2}. We have shown elsewhere \cite{mmw,mmpw} that this empirical distribution is very well fitted by a power law of the form

\begin{equation}
h(x)=\frac{(\beta-1)}{2\gamma(1+|x|/\gamma)^\beta},
\label{empiricalh}
\end{equation}
where $\beta=5.52$ and $\gamma=2.64\times 10^{-4}$. In this case the mean waiting time between jumps is $\langle\tau\rangle=23.65\ s$. The rest of parameters are
$$
\kappa=\frac{\gamma\sqrt{2}}{\sqrt{(\beta-2)(\beta-3)}}, 
$$
and
$$
H(0)=\frac{(\beta-1)}{\sqrt{2(\beta-2)(\beta-3)}}, \quad 
H'(0+)=-\frac{\beta(\beta-1)}{(\beta-2)(\beta-3)},
$$
with numerical values $\kappa=1.25 \times 10^{-4}$, $H(0)=1.07$ and $H'(0+)=-2.81$. 

\begin{figure}
\epsfig{file=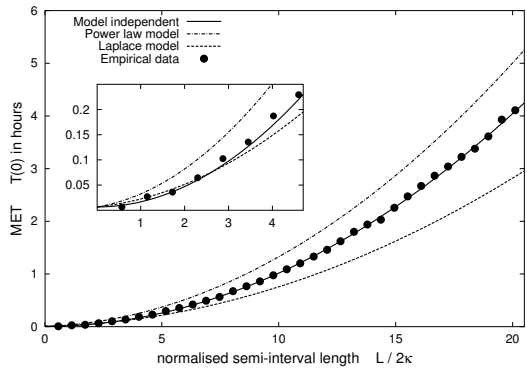, width=7cm}
\caption{Dots represent the empirical MET (measured in hours) as a function of the normalized semi-interval $L/2\kappa$, from high frequency data of the U.S. dollar/Deutsche mark futures market. We also plot several theoretical predictions explained in the text. The best adjustment corresponds to Eq. (\ref{metapproxsym}) with all parameters directly estimated from data without making any hypothesis on the form of the jump distribution $h(x)$.} 
\label{fig3}
\end{figure}

In Fig.~\ref{fig3} we compare the empirical MET from the U.S. dollar/Deutsche mark data with the analytical approximation given by Eq.~(\ref{metapproxsym}). We see there that when $h(x)$ is estimated by the power law (\ref{empiricalh}), with the numerical values for $\kappa$, $H(0)$, and $H'(0+)$ given above, the theoretical expression does not fit the empirical data in a satisfactory manner. The reason for it lies in the fact the approximation for $T(x_0)$ given by Eq.(\ref{metapproxsym}) basically depends on the values of $h(x)$ around $x=0$ while the power law density~(\ref{empiricalh}) has been obtained to fit the tails of the jump distribution rather than its center. 

On the other hand, we can easily evaluate the values of $\kappa$ and $H(0)$ directly from data without assuming any hypothesis on the form of the jump distribution. This yields $\kappa=1.70 \times 10^{-4}$, $H(0)=4.45 \times 10^{-3}$. We also need an estimation for $H'(0+)$ although this parameter is quite difficult to evaluate because data binning entails a certain degree of arbitrariness. Moreover the very existence of tick units questions the concept of derivative. A first approach to solve this problem would be to evaluate $H'(0+)$ using finite differences, which gives $H'(0+)=0.55$. Nevertheless if we represent $T(0)$ as given in Eq. (\ref{metapproxsym}) using the numerical values of $\kappa$, $H(0)$ and $H'(0+)$ just reported, then the resulting function fits to some extend the empirical MET albeit being not completely satisfactory. A better fit of $T(0)$ can be obtained by using the same values of $\kappa$ and $H(0)$ but estimating $H'(0+)$
in order to furnish an optimal fit for the MET which finally yields the value $H'(0+)=1.54$. This is represented by the solid line curve of Fig. \ref{fig3}. Note that this final estimation shows that the approximate expressions for the MET  given by Eqs.~(\ref{metapprox}) and~(\ref{metapproxsym}) are very efficient since they almost exactly reproduce the empirical behavior of the MET far away from the validness of the approximation itself, which we recall was only effective for small values of the normalized interval $L/\kappa$. We can thus conclude that the MET presents a quadratic nature regardless the length of the interval. All of this seems to indicate that the approximations given by Eqs. (\ref{metapprox}) and (\ref{metapproxsym}) have a more universal character than one might expect in advance. 

We finally observe that the Laplace density given by Eq. (\ref{laplace}), which yields the exact MET (\ref{exactmetsim}), results in a good approximation of the MET for small values of the interval as we can also see in Fig. \ref{fig3}.In this regard the Laplace density can be considered as a first approximation to any more realistic jump distribution. 

\section{Conclusions}

Using the CTRW framework for market microstructure, we have developed a somewhat little known aspect of the problem: the study of extreme events, specially the mean exit time out of an interval, $T(x_0)$, and the mean first-passage time to some critical value, $T_c(x_0)$, although the latter turns out to be infinite in many situations. We have shown that these extreme times obey an integral equation which depends on the jump distribution $h(x)$ and the mean waiting time $\langle\tau\rangle$. We have exactly solved the integral equation for the MET in the case where the jump distribution is governed by a Laplace (symmetric exponential) probability density function. We have compared this MET with that of the ordinary random walk model and showed that the MET for the Laplace density is bigger than $T_{\mbox{\tiny RW}}$ (the MET when $X(t)$ is assumed to be the Wiener process). This seems to indicate that the models based on the Wiener process may underestimate the MET (specially for small intervals). In other words, RW models imply that the return process escapes faster from a give interval than models based on the CTRW. We believe that this can have practical consequences in risk control as well as in pricing exotic derivatives. 

We have also solved the integral equation for the MET using an approximate scheme which yields a solution valid for a general $h(x)$ but when the length $L$ of the interval is small (in an appropriate dimensionless scale). We have showed that for the Laplace form of $h(x)$ the approximate solution for small $L$ coincides with the exact solution, therefore showing that the Laplace is a ``first-order approximation" to any sufficiently smooth and symmetrical jump distribution. We have applied the approximate solution to real data (high frequency data on U.S. dollar-Deutsche mark futures market) with a very good agreement between the empirical and the theoretical MET. 

We finally mention that more complete measures of the extreme events, such as the survival probability, with greater impact on risk control are under present investigation and some results will be published soon.

\begin{acknowledgments}
This work has been supported in part by Direcci\'on General de
Investigaci\'on under contract No. BFM2003-04574 and by Generalitat de Catalunya under contract No. 2001 SGR-00061.
\end{acknowledgments}


\begin{thebibliography}{999}
\bibitem{footnote1} Note that by definition ({\it cf} Eq. (\ref{x})) $x_0=0$. Nevertheless for the clarity of the exposition we keep a general value for $x_0$, although in practical applications we will set $x_0=0$.
\bibitem{footnote2} If there is no confusion and in order to keep notation simple, we will drop the subscript and write $T(x_0)$ for the MET out of $[a,b]$.
\bibitem{footnote3} In proving this one has to take into account that $h(x)$ is nonnegative on the entire real line.
\end{thebibliography}
\end{document}